\documentclass[prl,twocolumn,showpacs,preprintnumbers,english,amsmath,amssymb]{revtex4}

\usepackage{graphicx,epsf,here}
\usepackage{dcolumn}
\usepackage{bm}

\begin{document}

\preprint{}

\title{Ballistic and diffusive: theory of vortices in the two-band superconductor MgB$_2$}
\author{K. Tanaka$^{1,2}$, D. F. Agterberg$^{1,3}$, J. Kopu$^{4,5}$,
and M. Eschrig$^{5}$} 
\affiliation{$^{1}$Argonne National Laboratory, Argonne, IL 60439, U.S.A. \\
$^{2}$Department of Physics and Engineering Physics, University of Saskatchewan, 
Saskatoon, SK, Canada S7N 5E2 \\
$^{3}$Department of Physics, University of Wisconsin -
Milwaukee, P.O. Box 413, Milwaukee, WI 53211, U.S.A\\
$^{4}$Low Temperature Laboratory, 
Helsinki University of Technology, PO Box 2200, FIN-02015 HUT, Finland \\
$^{5}$Institut f\"ur Theoretische Festk\"orperphysik,
Universit\"at Karlsruhe, 76128 Karlsruhe, Germany}
\date{December 6, 2005}
\begin{abstract}
Motivated by the recent results on impurity effects in MgB$_2$,
we present a theoretical model for a two-band superconductor in which the 
character of quasiparticle motion is ballistic in one band 
and diffusive in the other. 
We apply our model to calculate the
electronic structure in the vicinity of an isolated vortex. We assume
that superconductivity in the diffusive ($\pi$) band is induced by
that in the clean ($\sigma$) band, as suggested by experimental evidence for MgB$_2$.
We focus our attention to the spatial variations of the order parameter,
the current density, and the vortex core spectrum 
in the two bands. 
Our results indicate that the coupling to the $\pi$ band 
can lead to the appearance of additional bound states near the gap edge in
the $\sigma $ band that are absent in the single-band case.

\end{abstract}

\maketitle

It is now well established that $\rm MgB_2$ is a
two-gap superconductor, and its essential superconducting properties
are well described by an isotropic $s$-wave two-band model
\cite{nagamatsu01,kwok03,shulga01,liu01,bouguet01,schmidt02,iavarone02}.
The two `bands' in $\rm MgB_2$ are the `strong' $\sigma$ band that
arises from the boron $\sigma $-orbitals
(with the energy gap $\Delta_\sigma\approx 7.2 {\rm meV}$), 
and the `weak' $\pi$ band that
derives from the boron $\pi $ orbitals
($\Delta_\pi\approx 2.3 {\rm meV}$). 
Despite the fact that each `band' consists in reality of a pair of bands,
the variation of the gap within each corresponding pair of Fermi-surface sheets
can be neglected
\cite{choi02,mazin04}.
The two energy gaps are observed to vanish at a common
transition temperature $T_c$: no second transition has been observed, and
there is evidence of induced superconductivity in the $\pi$ band
\cite{schmidt02,eskildsen02,geerk05}.

There have been considerable efforts to understand impurity effects in $\rm
MgB_2$ \cite{kwok03}. Besides the potential applications of $\rm
MgB_2$ as magnetic devices \cite{gurevich03}, these studies have aimed
at testing a prediction of the two-band model: the reduction of $T_c$ and the
gap ratio by non-magnetic impurities. However, Mazin {\it et al.}
\cite{mazin02,erwin03} have shown that this does not apply to $\rm
MgB_2$, i.e. to Mg-site impurities or defects, which are more
favourable energetically than those at B sites. While the $\pi$ band
is strongly affected by such impurities, the $\sigma$ band
is more robust, and there is little mixing of the two bands because of
negligible interband scattering \cite{mazin02,erwin03}. 
This is consistent with accumulating
experimental evidence suggesting that the $\sigma$ and $\pi$ bands
are essentially in the ballistic and diffusive limit, respectively
\cite{clean_sigma}
(see also references in Refs.~\cite{mazin02,erwin03}).
Nevertheless, so far in theoretical studies both of the two bands have been
assumed to be either in the clean \cite{nakai02} 
or in the dirty limit \cite{koshelev03}.

In this Letter, we examine theoretically the effects of induced
superconductivity and impurities on the vortex-core structure in a
two-band superconductor. Our work is motivated by the experimental
investigation of the vortex state in $\rm MgB_2$ using scanning
tunnelling spectroscopy \cite{eskildsen02}.
We present a novel formulation
of coupled quasiclassical Eilenberger and Usadel equations to describe
a multiband superconductor with one ballistic and one diffusive
band with negligible interband scattering:
the two bands are assumed to be coupled only by the pairing interaction. 

We apply our model to calculate numerically the local density of
states (LDOS) and supercurrent density around an isolated vortex. We
examine in detail the intriguing spatial variation of these
quantities and the order parameter in the two bands.
A particularly interesting result emerging from our studies is the
possibility of additional bound states near the gap edge in the
$\sigma$ band.

Our model is based on the
equilibrium quasiclassical theory of superconductivity, where the
physical information is contained in the Green function, or
propagator, $\hat{g}(\epsilon, {\bf p}_{F\alpha}, {\bf R})$. Here
$\epsilon$ is the quasiparticle energy measured from the chemical
potential, ${\bf p}_{F\alpha}$ the quasiparticle momentum on the Fermi
surface of band $\alpha \in \{\sigma,
\pi\}$, and ${\bf R}$ is the spatial coordinate (the hat refers
to the 2$\times$2 matrix structure of the propagator in the
particle-hole space). In the clean $\sigma$ band,
$\hat g_\sigma(\epsilon, {\bf
p}_{F\sigma}, {\bf R})$ satisfies the Eilenberger equation
\cite{eilenberger}
\begin{equation}
\left[ \epsilon \hat \tau_3 - \hat \Delta_\sigma 
,\; \hat g_\sigma \right]
+i {\bf v}_{F\sigma} \cdot {\nabla }  \hat g_\sigma
= \hat 0,
\label{eil}
\end{equation}
where ${\bf v}_{F\sigma}$ is the Fermi velocity and $\hat \Delta_\sigma$
the (spatially varying) order parameter
in the $\sigma$ band. Throughout this work, we
ignore the external magnetic field (this is justified because
MgB$_2$ is a strong type-II superconductor).
The coherence length in the $\sigma $ band is defined as $\xi_{\sigma
}=v_{F\sigma }/ 2\pi T_c$.

In the presence of strong impurity scattering, the Green function has
no momentum dependence and the Eilenberger equation reduces to the
Usadel equation \cite{usadel70}. We assume this to be the appropriate
description for the dirty $\pi$ band, and take the
propagator $\hat g_\pi(\epsilon,{\bf R})$ to satisfy
\begin{equation}
\left[ \epsilon \hat \tau_3 - \hat \Delta_\pi ,\; \hat g_\pi \right] 
+ \frac{D}{\pi} {\nabla } \cdot (\hat g_\pi {\nabla } \hat g_\pi )
= \hat 0.
\end{equation}
The diffusion constant $D$ defines the $\pi$-band
coherence length as $\xi_{\pi
}=\sqrt{D/2\pi T_c}$. 
Additionally, both propagators are normalized according to
$\hat g_\sigma^2=\hat g_\pi^2=-\pi^2 \hat 1$.

We assume that the quasiparticles in different bands are coupled only
through the pairing interaction, neglecting interband scattering
by the impurities. 
The gap equations for the multiband system are given as
\begin{equation}
\Delta_\alpha ({\bf R}) = \sum_{\beta} V_{\alpha\beta} N_{F\beta}
{\cal F}_\beta ({\bf R}),
\label{gapeq}
\end{equation}
where $\alpha,\beta \in \{\sigma,\pi\}$,
$\hat\Delta_\alpha=\hat\tau_1~\rm{Re}~\Delta_\alpha-\hat\tau_2~\rm{Im}~\Delta_\alpha$,
the coupling matrix $V_{\alpha\beta}$ determines the strength of the
pairing interaction, $N_{F\beta}$ is the Fermi-surface density of
states on band $\beta$, and
\begin{eqnarray}
{\cal F}_{\sigma }({\bf R})&\equiv&
\int_{-\epsilon_c}^{\epsilon_c} {d\epsilon\over 2 \pi i}\,
\langle f_\sigma (\epsilon , {\bf p}_{F\sigma},{\bf R}) \rangle_{{\bf p}_{F\sigma}}
\, {\rm tanh}\left({\epsilon\over 2 T}\right),
\label{F1} \nonumber \\
{\cal F}_{\pi }({\bf R})&\equiv&
\int_{-\epsilon_c}^{\epsilon_c} {d\epsilon\over 2 \pi i}\,
f_\pi (\epsilon, {\bf R} )
\, {\rm tanh}\left({\epsilon\over 2 T}\right).
\label{F2}
\end{eqnarray}
Here $f_\alpha$ is the upper off-diagonal (1,2) element of the matrix
propagator $\hat g_\alpha$, $\langle\cdots\rangle_{{\bf p}_{F\sigma}}$
denotes averaging over the $\sigma$ band Fermi surface, and
$\epsilon_c$ is a cut-off energy.

We solve the system of equations (\ref{eil})--(\ref{F2}) numerically.
The normalization condition is taken into account with the Riccati
parameterisation for the Green functions \cite{schopohl,matthias1}. After
self-consistency has been achieved for the order parameter, the (for
the $\sigma$ band angle-resolved) LDOS in each band can be calculated
as
\begin{eqnarray}
N_{\sigma }(\epsilon,{\bf p}_{F\sigma}, {\bf R})/N_{F\sigma}&=&-
~{\rm Im}~g_\sigma (\epsilon , {\bf p}_{F\sigma},{\bf R}) /\pi
,
\nonumber \\
N_{\pi }(\epsilon,{\bf R})/N_{F\pi}&=&-
~{\rm Im}~g_\pi (\epsilon,{\bf R})/\pi,
\label{LDOS}
\end{eqnarray}
where $g_\alpha$ is the upper diagonal (1,1) element of $\hat g_\alpha$.

The current density around the vortex has contributions from
both the $\pi$ band and the $\sigma $ band. The corresponding expressions are
($e=-|e|$ is the electron charge)
\begin{eqnarray}
\frac{{\bf j}_{\sigma }({\bf R})}{2eN_{F\sigma }}&=&
\int_{-\infty}^{\infty} {d\epsilon\over 2\pi }\,
\langle {\bf v}_{F\sigma } {\rm Im}~g_\sigma \rangle_{{\bf p}_{F\sigma}}
\tanh\left( \frac{\epsilon}{2T}\right), \; \nonumber
\\
\frac{{\bf j}_{\pi}({\bf R})}{2eN_{F\pi}}&=&
\frac{D}{\pi} \int_{-\infty}^{\infty} {d\epsilon\over 2\pi }\,
{\rm Im}~[f_\pi^{\ast }  {\nabla } f_\pi]
\tanh\left( \frac{\epsilon}{2T}\right). \;
\label{CD}
\end{eqnarray}

Throughout this work,
we focus on the effects of purely induced superconductivity in the $\pi$ band.
Diagonalization of the coupling matrix in
Eq.~(\ref{gapeq}) decouples the gap equations.
The larger of the two eigenvalues of the matrix $V_{\alpha \beta }N_{F\beta }$,
denoted by $\lambda^{(0)}$, determines the critical temperature
$T_c$ and can be eliminated together with $\epsilon_c$.
The smaller eigenvalue $\lambda^{(1)}$ is parameterized by the cut-off independent
combination $\Lambda= \lambda^{(0)}\lambda^{(1)}/ (\lambda^{(0)}-\lambda^{(1)})$.
The pairing interactions have been calculated with ab initio methods,
employing an electron-phonon coupling model \cite{liu01,choi02}; 
from these studies, taking Coulomb repulsion into
account \cite{mazin04}, we estimate $\Lambda <0.3 $ for MgB$_2$. We
present results for $\Lambda =-0.1$ (implying a weak repulsion in the
subdominant $\lambda^{(1)}$ channel).
The ratio of the
magnitude of the bulk gaps $\rho =|\Delta_\pi^{\rm
bulk}|/|\Delta_\sigma^{\rm bulk}|$ near $T_c$ parameterizes the strength
of the induced superconductivity in the $\pi $ band; experimentally
$\rho \approx 0.3$ in MgB$_2$.
For simplicity, we have set $N_{F\sigma}=N_{F\pi}$
in our calculations; the densities of states of the
two bands are indeed of comparable size \cite{choi02,mazin04}.
A cylindrical Fermi surface was used for the two-dimensional $\sigma$
band. 
For the ratio of the coherence lengths in the two bands we present
results for $\xi_\pi/\xi_\sigma$=1, 3, and 5. 
An estimate from experiments \cite{eskildsen02,serventi04}
gives for MgB$_2$ a value between 1 and 3.

\begin{figure}
\begin{minipage}{\columnwidth}
\includegraphics[width=0.48\columnwidth]{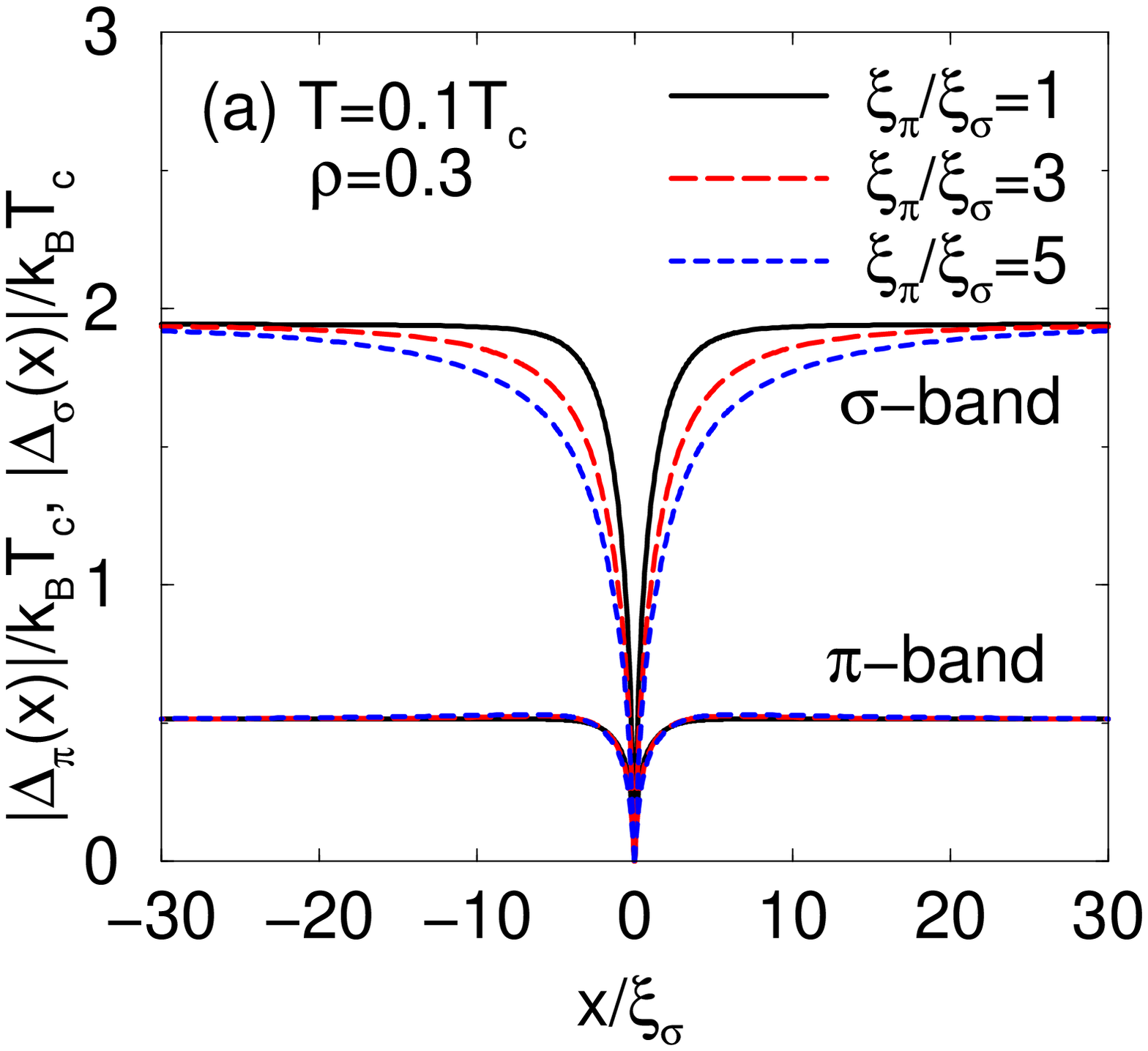}
\hfill
\includegraphics[width=0.48\columnwidth]{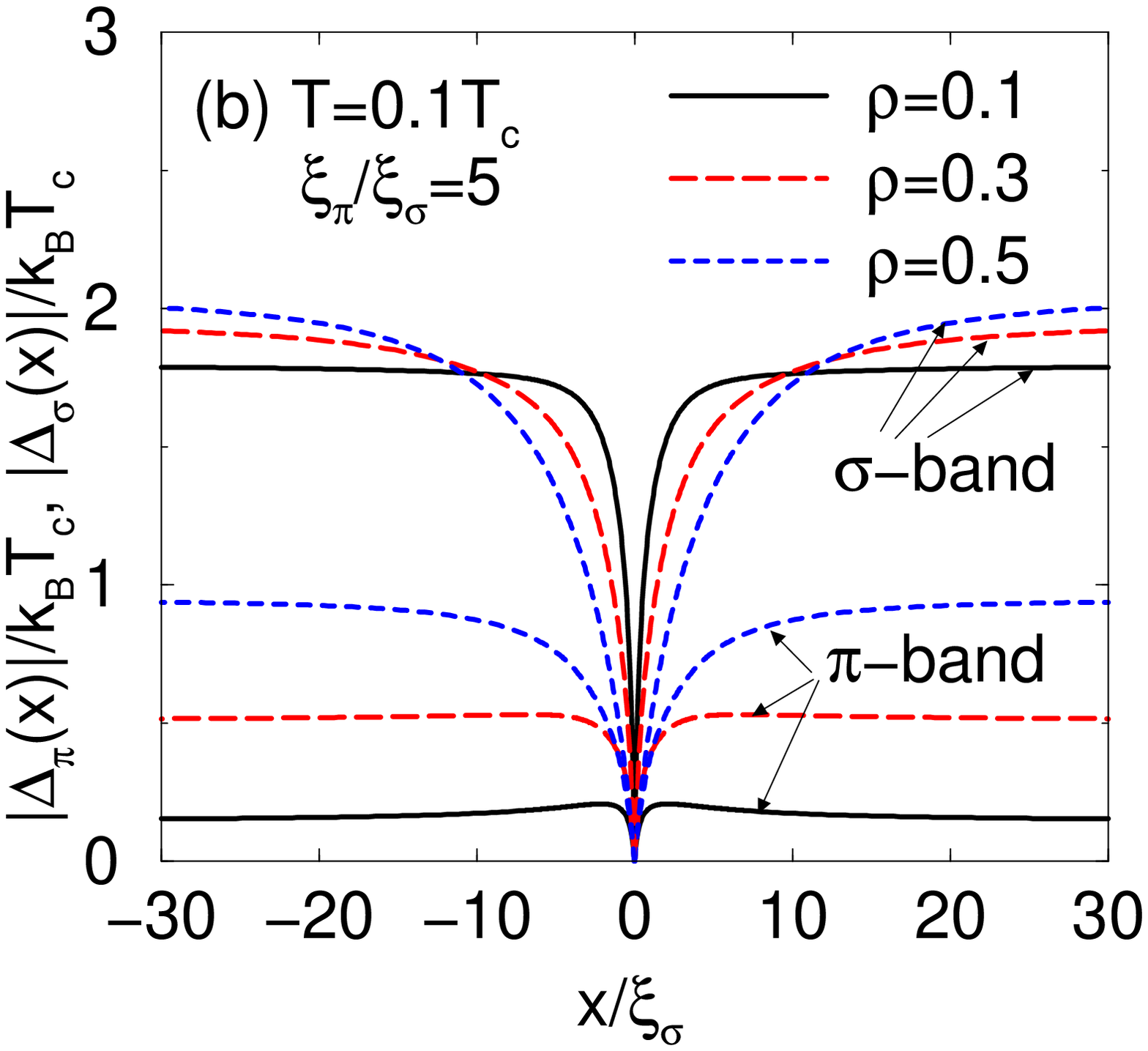}
\end{minipage}
\caption{
Magnitude of the order parameter, $|\Delta_{\sigma,\pi} (x)|$,
in the $\sigma$- and $\pi $ band, as a function of coordinate $x$ on
a path through the vortex center, at $T=0.1 T_c$:
(a) for different ratios $\xi_\pi/\xi_\sigma $ and fixed
strength of the induced $\pi$-band gap, parameterized by the mixing ratio (see text), 
$\rho =0.3$;
(b) for different $\rho$ and fixed $\xi_\pi/\xi_\sigma=5$.
A weak effective repulsion due to Coulomb interaction was assumed 
in the subdominant $\lambda^{(1)}$ pairing channel,
parameterized by $\Lambda=-0.1$ (see text).
}
\label{fig1}
\end{figure}
In Fig.~\ref{fig1} we present the order-parameter magnitudes for each
band as a function of coordinate $x$ along a path through the vortex
center. In Fig.~\ref{fig1} (a) we show the order parameter variation at $T=0.1 T_c$
for several coherence-length ratios
$\xi_\pi/\xi_\sigma $ and gap ratio $\rho=0.3$. Surprisingly, we find that an increase of $\xi_\pi$
results in an increase of the recovery length of the order parameter 
(the characteristic length
over which the order parameter changes from zero to the bulk value) in
the $\sigma$ band, while the recovery length in the $\pi$ band is barely affected. 
Thus, the recovery lengths in the two bands
can differ considerably, even though the superconductivity in
the $\pi $ band is induced by the $\sigma $ band. 
In Fig.~\ref{fig1} (b), the gap ratio $\rho $ is varied for fixed
$\xi_\pi/\xi_\sigma=5$. One can see that,
apart from the well-known increase of the bulk $\Delta_\sigma/T_c$ ratio
with increasing bulk $\Delta_\pi$,
the recovery length in the $\sigma $ band increases considerably with increasing $\rho $.
The $\pi$-band recovery length is also enhanced together with that of the $\sigma$ band
if $\rho >0.3$, as clearly seen for $\rho=0.5$. 
For $\rho=0.1$, $\Delta_\pi$ near the vortex core is enhanced from its
bulk value.

\begin{figure}
\begin{minipage}{\columnwidth}
\includegraphics[width=0.48\columnwidth]{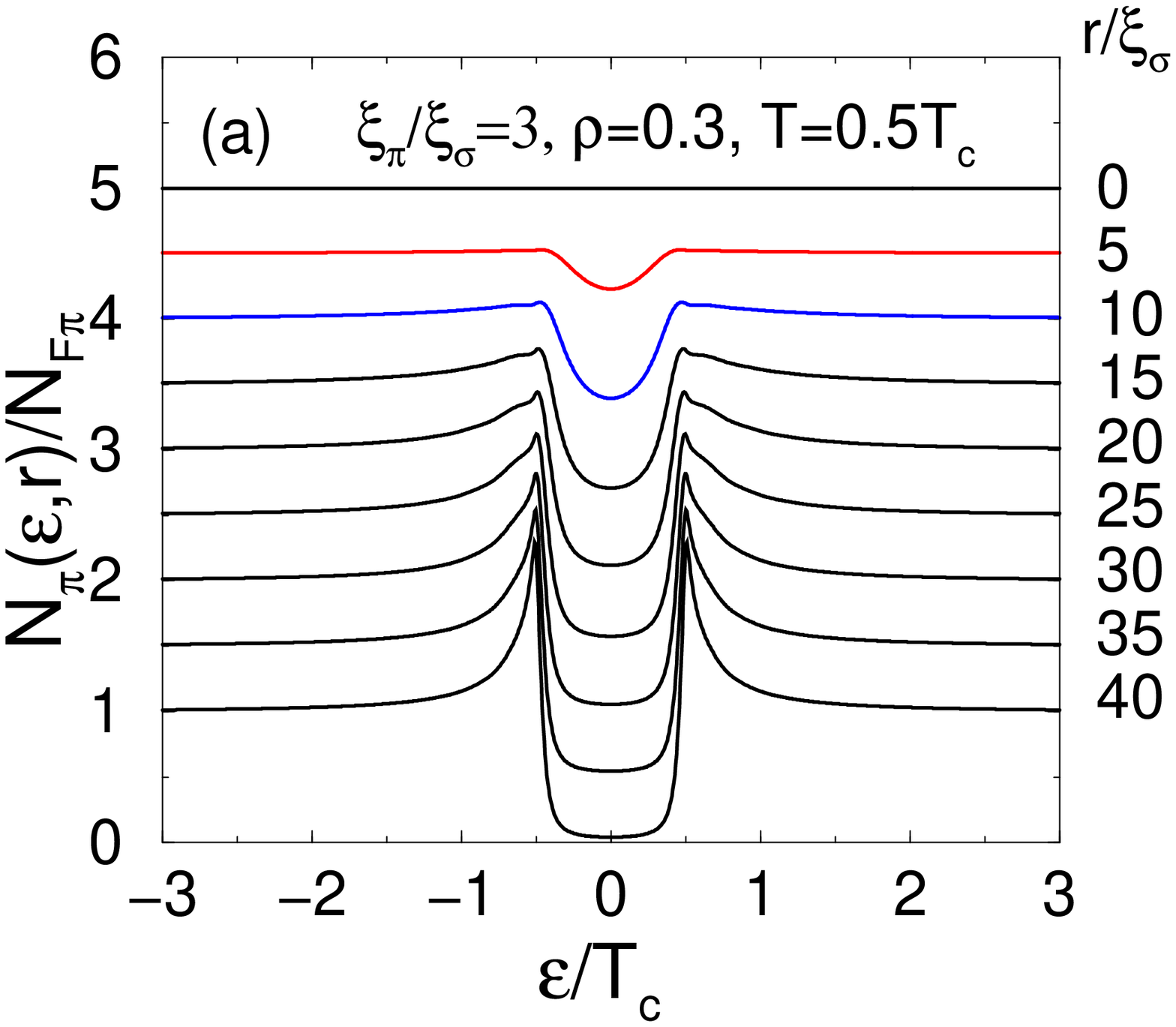}
\hfill
\includegraphics[width=0.48\columnwidth]{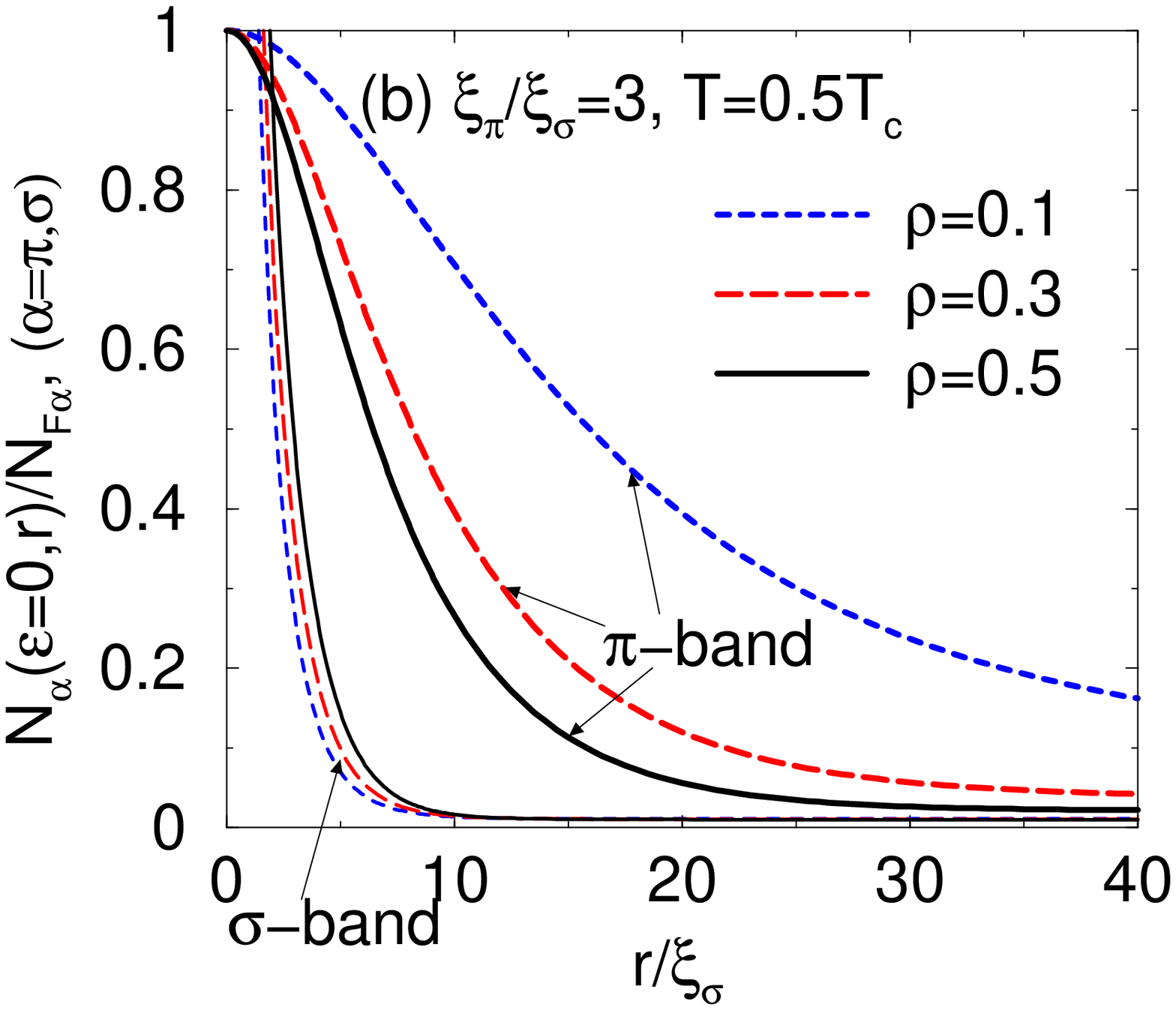}
\end{minipage}
\caption{
(a) Local density of states (LDOS) for the $\pi $ band,
as a function of energy $\epsilon $ for
different distances $r$ from the vortex center.
The spectra are shifted in vertical direction for convenience.
(b) LDOS at the chemical potential, $\epsilon=0$,
as a function of $r$ for different $\rho$
and fixed $T=0.5T_c$, $\xi_\pi/\xi_\sigma =3$.
}
\label{fig3}
\end{figure}

To understand these effects we have performed calculations
for different values of $\Lambda $ that characterizes the
Coulomb repulsion in the $\pi$ band. We have found that for $\Lambda >0$
the ratio $|\Delta_\pi|/|\Delta_\sigma |$ is reduced in the
core region with respect to its bulk value, however is enhanced for 
$\Lambda <0$. This leads to a renormalization of the
recovery lengths of the order parameters in both bands.

\begin{figure}
\begin{minipage}{\columnwidth}
\includegraphics[width=0.5\columnwidth]{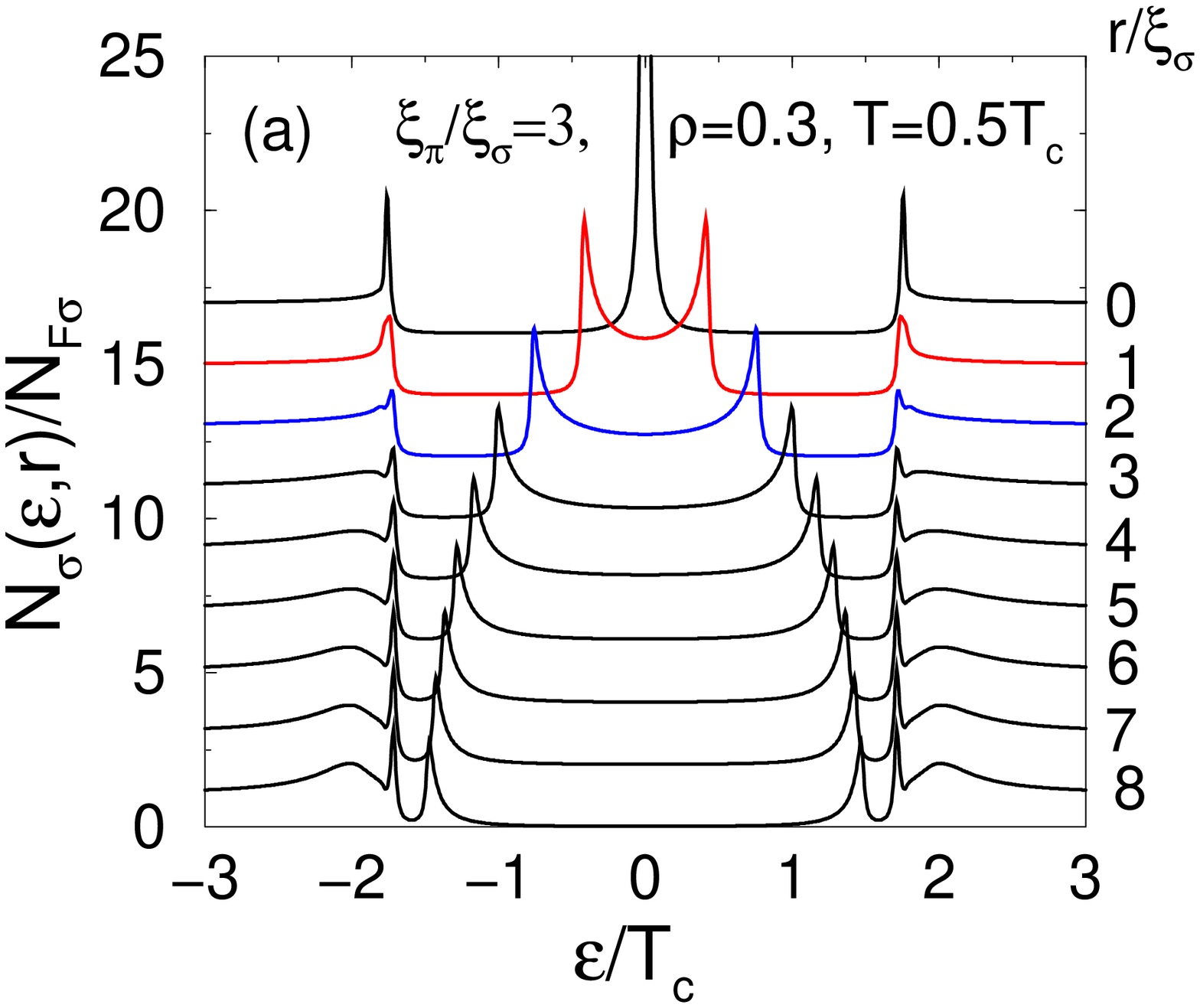}
\hfill
\includegraphics[width=0.45\columnwidth]{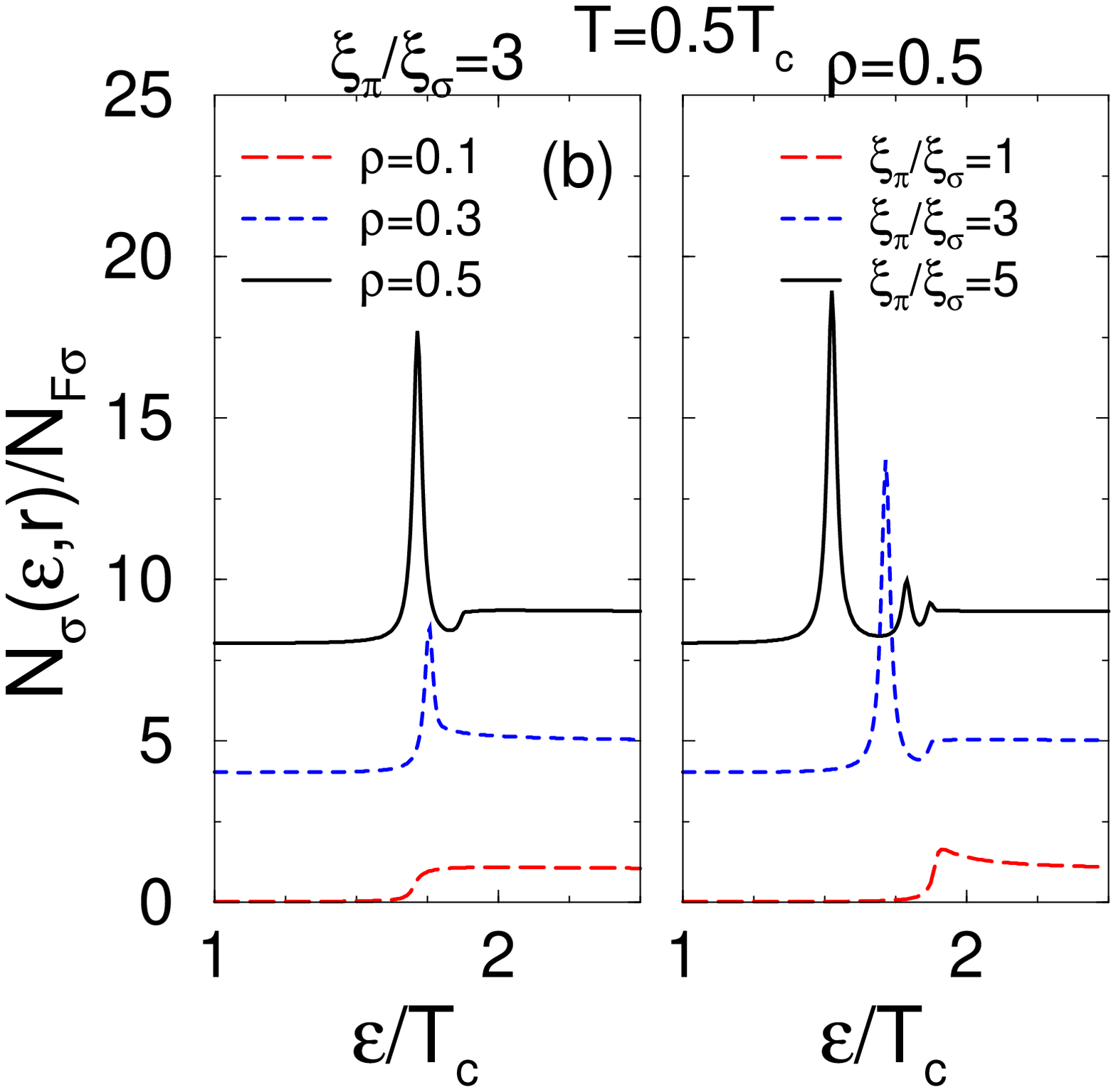}
\end{minipage}\\
\begin{minipage}{\columnwidth}
\includegraphics[width=0.5\columnwidth]{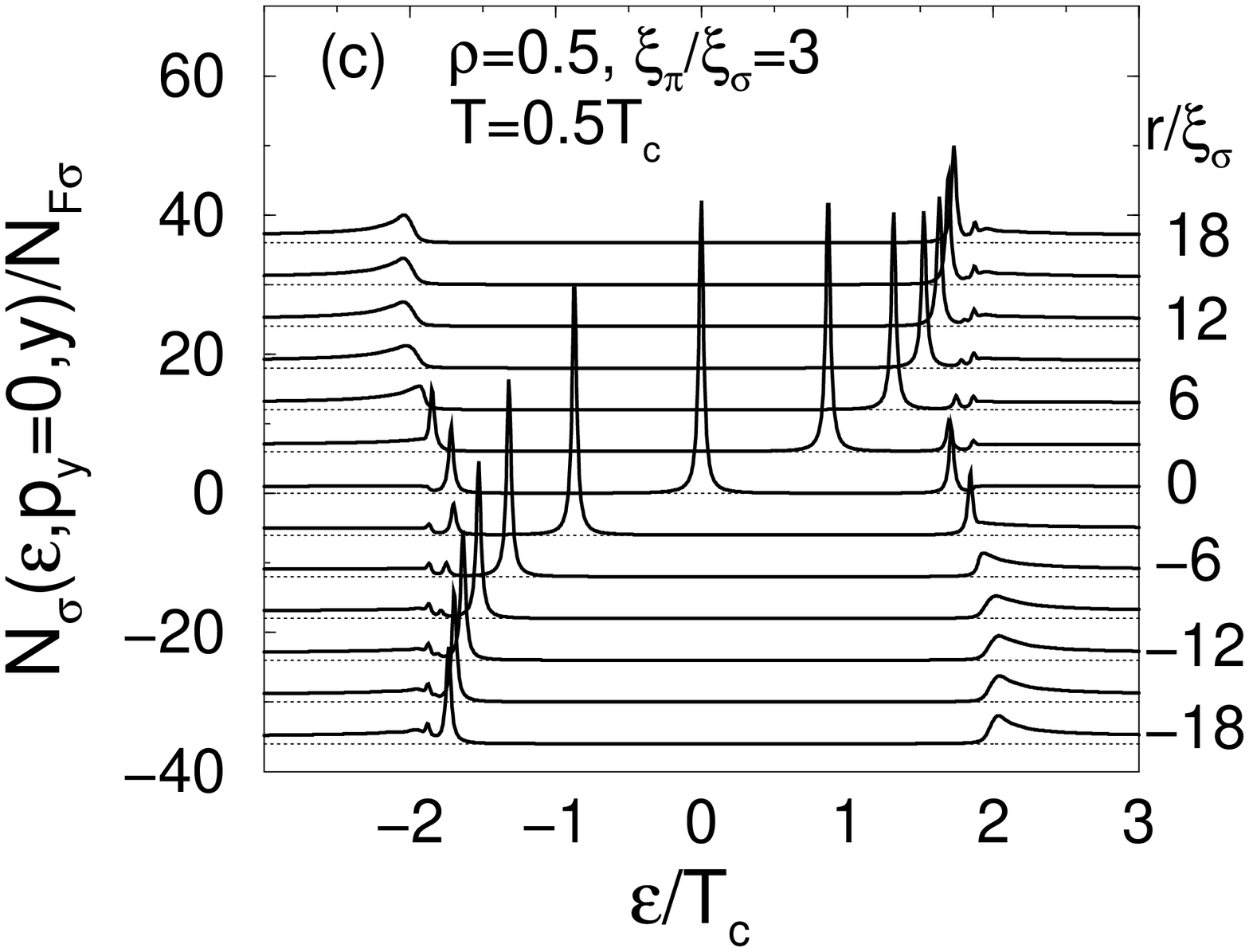}
\hfill
\includegraphics[width=0.48\columnwidth]{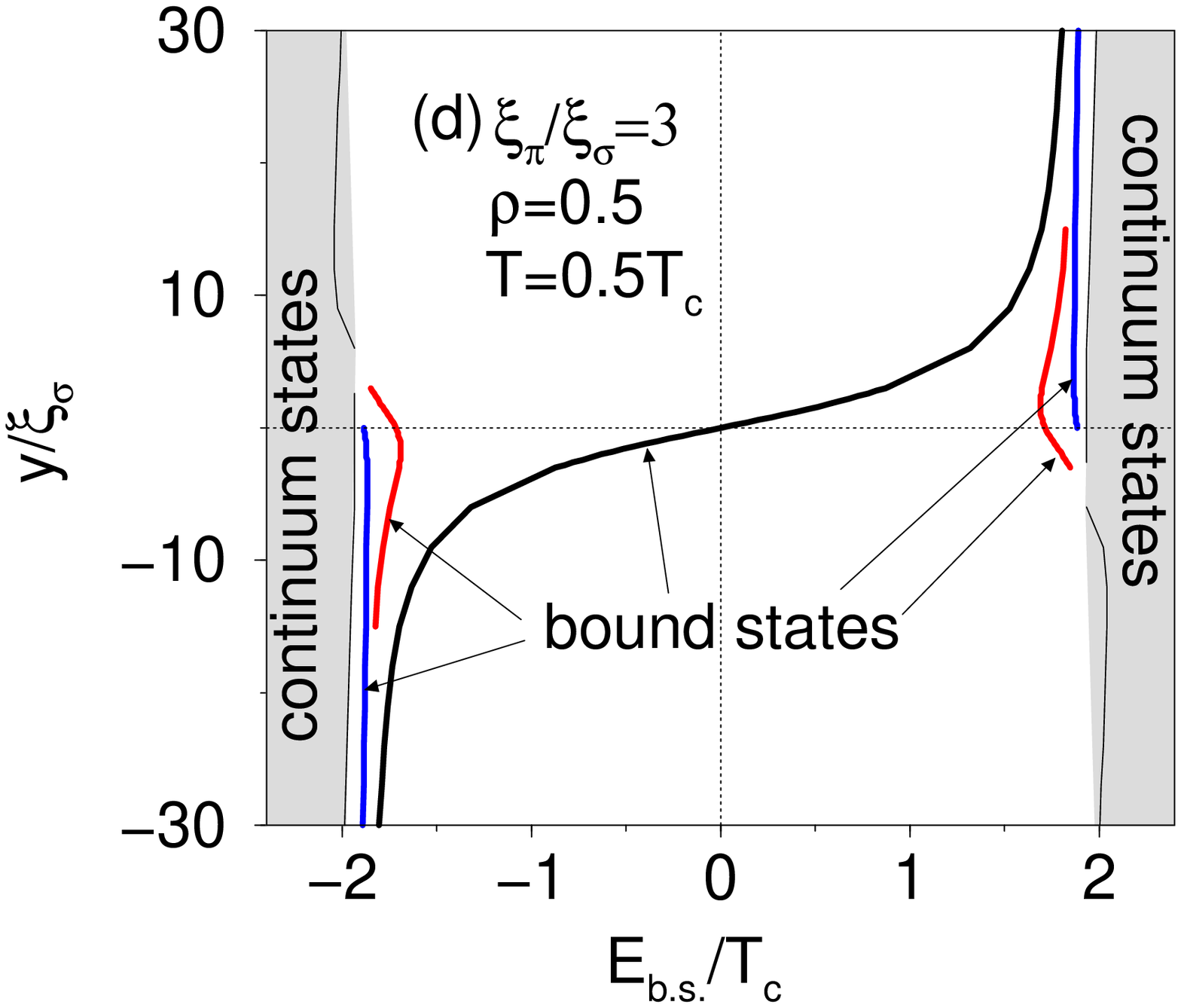}
\end{minipage}
\caption{
(a) LDOS for the $\sigma $ band
as a function of energy $\epsilon $ for
different distances $r$ from the vortex center.
(b) The development of an extra bound state near the gap edge in the
$\sigma $ band for different parameter combinations. The bound state
develops for sufficiently large mixing ratio $\rho $ and coherence length
ratio $\xi_\pi/\xi_\sigma $.
(c) Angle-resolved LDOS
for quasiparticles with momentum along the $x$ axis at positions along
the $y$ axis in the $\sigma $ band, showing the bound states
as a function of $y$. The dispersion of the bound states as a function
of $y$ is also shown in (d) for $\rho=0.5$, 
$\xi_\pi/\xi_\sigma =3$ and $T=0.5T_c$.
Spectra in (a)-(c) are shifted in vertical direction for convenience.
}
\label{fig2}
\end{figure}

In Fig.~\ref{fig3} we show the spectral properties of the $\pi$
band. The LDOS, shown in Fig.~\ref{fig3} (a), is flat at the vortex
center ($r=0$), in agreement with the experiment of
Ref.~\cite{eskildsen02}. Outside the vortex core the BCS density of
states is recovered. The decay of the zero-bias LDOS as a function of
radial coordinate is shown in Fig.~\ref{fig3} (b), and compared with
that of the $\sigma$ band [obtained from the data in Fig.~\ref{fig2} (a)]. The
decay length of the zero-bias DOS is clearly different for the two
bands. For the $\pi$ band it is given by $\xi_\pi \sqrt{\Delta_\sigma
/\Delta_\pi }$, and thus dominated by the parameter $\rho$. In the
$\sigma $ band the length scale of the decay is $\xi_\sigma$, and thus
shorter than that in the $\pi$ band. The existence of two apparent
length scales in the LDOS was also reported in the case of two clean
bands \cite{nakai02} and two dirty bands \cite{koshelev03}.

In Fig.~\ref{fig2} we show the vortex-core spectra in the $\sigma $
band.  
The LDOS,
$ N_{\sigma }(\epsilon,{\bf R})= 
\langle N_{\sigma }(\epsilon,{\bf p}_{F\sigma}, {\bf R}) \rangle_{{\bf p}_{F\sigma}} $,
as a function of energy for different distances from
the vortex center is plotted in Fig.~\ref{fig2} (a), showing the
well-known Caroli-de Gennes-Matricon bound-state bands at low
energies.  The new feature of our model is the additional bound states
in the vortex-core region near the gap edges, that are clearly visible
in Fig.~\ref{fig2} (a).  This is in strong contrast with the case of a
single-band superconductor, where the spectrum near the vortex center
is suppressed at the gap edges, showing neither a coherence peak nor
additional bound states.  
We note that the self-consistency
of the order-parameter profile is essential for discussing the
presence of the bound states at the gap edges.
We illustrate the development of these
additional bound
states in terms of the spectrum at the vortex center.
We find that for a given $\xi_\pi/\xi_\sigma $, the bound state
exists for $\rho$ larger than a certain critical value, as seen
in the left panel of Fig.~\ref{fig2} (b).
For a given $\rho$, on the other hand, the bound state develops if
$\xi_\pi/\xi_\sigma $ exceeds a critical value,
which is between 1 and 3 for $\rho=0.3$, as can be inferred from the
right panel of Fig.~\ref{fig2} (b).
The bound states are, e.g., clearly
resolved for $\rho=0.3$ and $\xi_\pi/\xi_\sigma =3$, values appropriate
for MgB$_2$. We have found a similar bound state spectrum also for
$\Lambda >0 $. The bound states move with the gap edge as a function of
temperature. 

\begin{figure}
\begin{minipage}{\columnwidth}
\includegraphics[width=0.48\columnwidth]{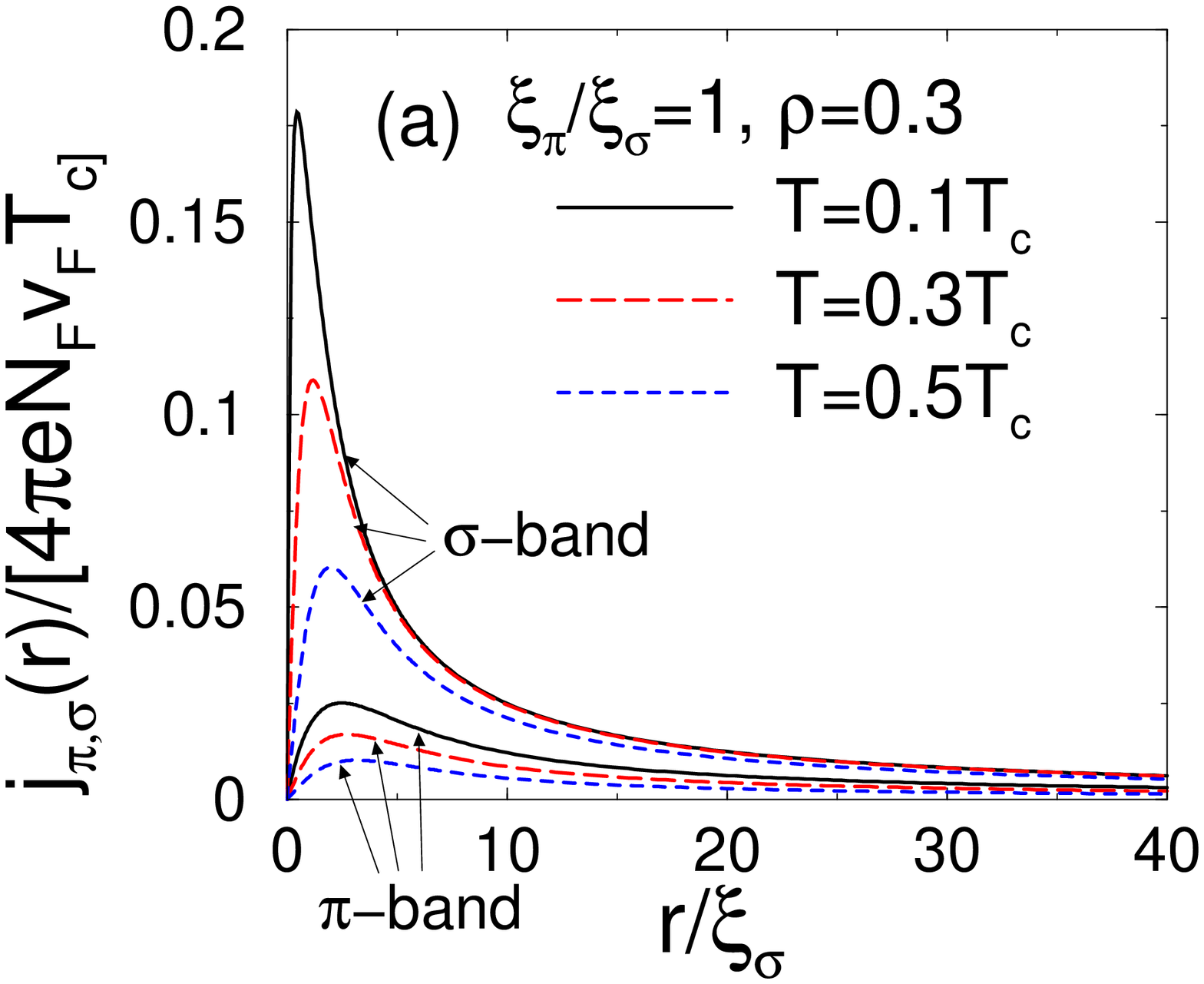}
\hfill
\includegraphics[width=0.48\columnwidth]{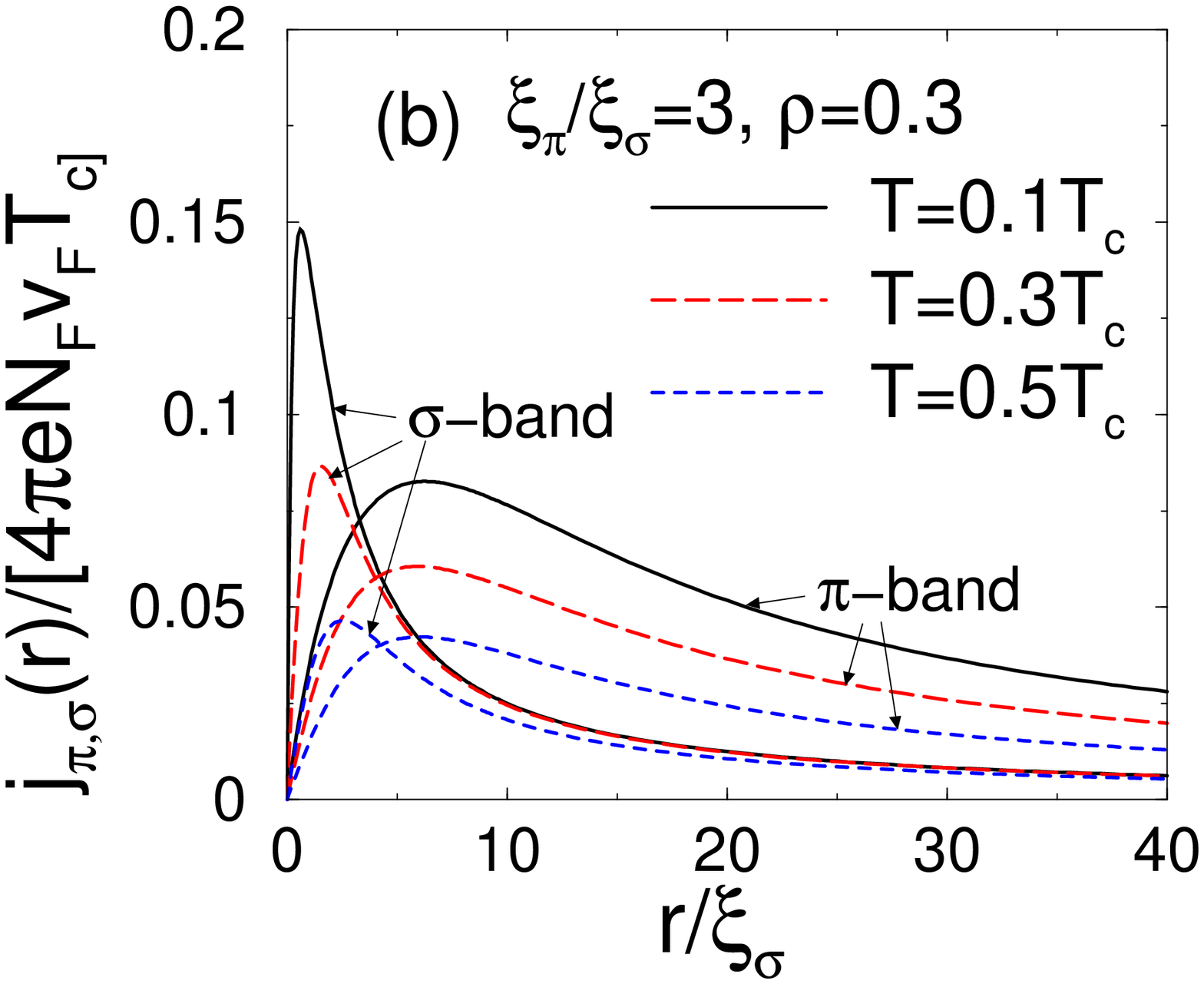}
\end{minipage}\\
\begin{minipage}{\columnwidth}
\includegraphics[width=0.48\columnwidth]{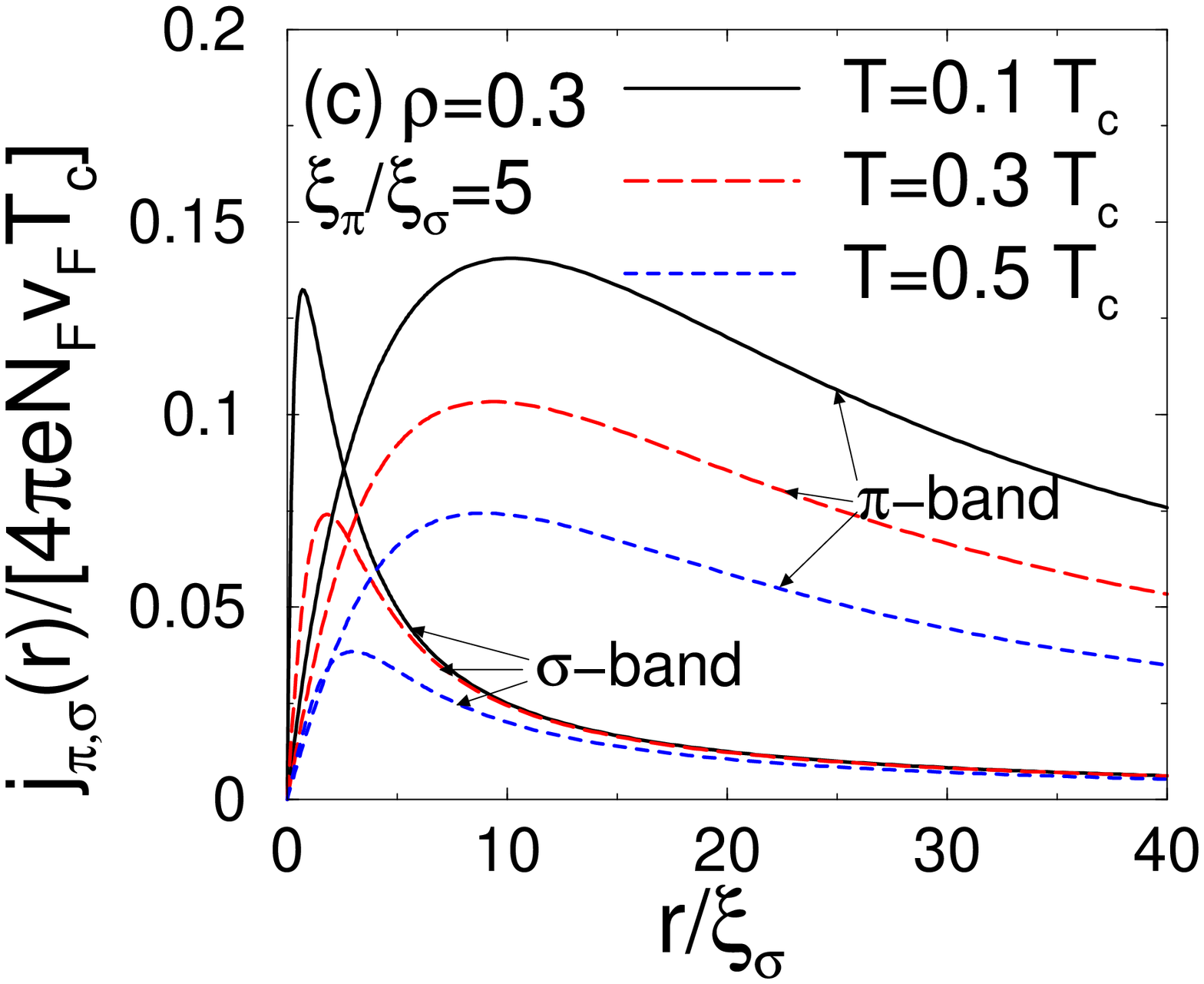}
\hfill
\includegraphics[width=0.48\columnwidth]{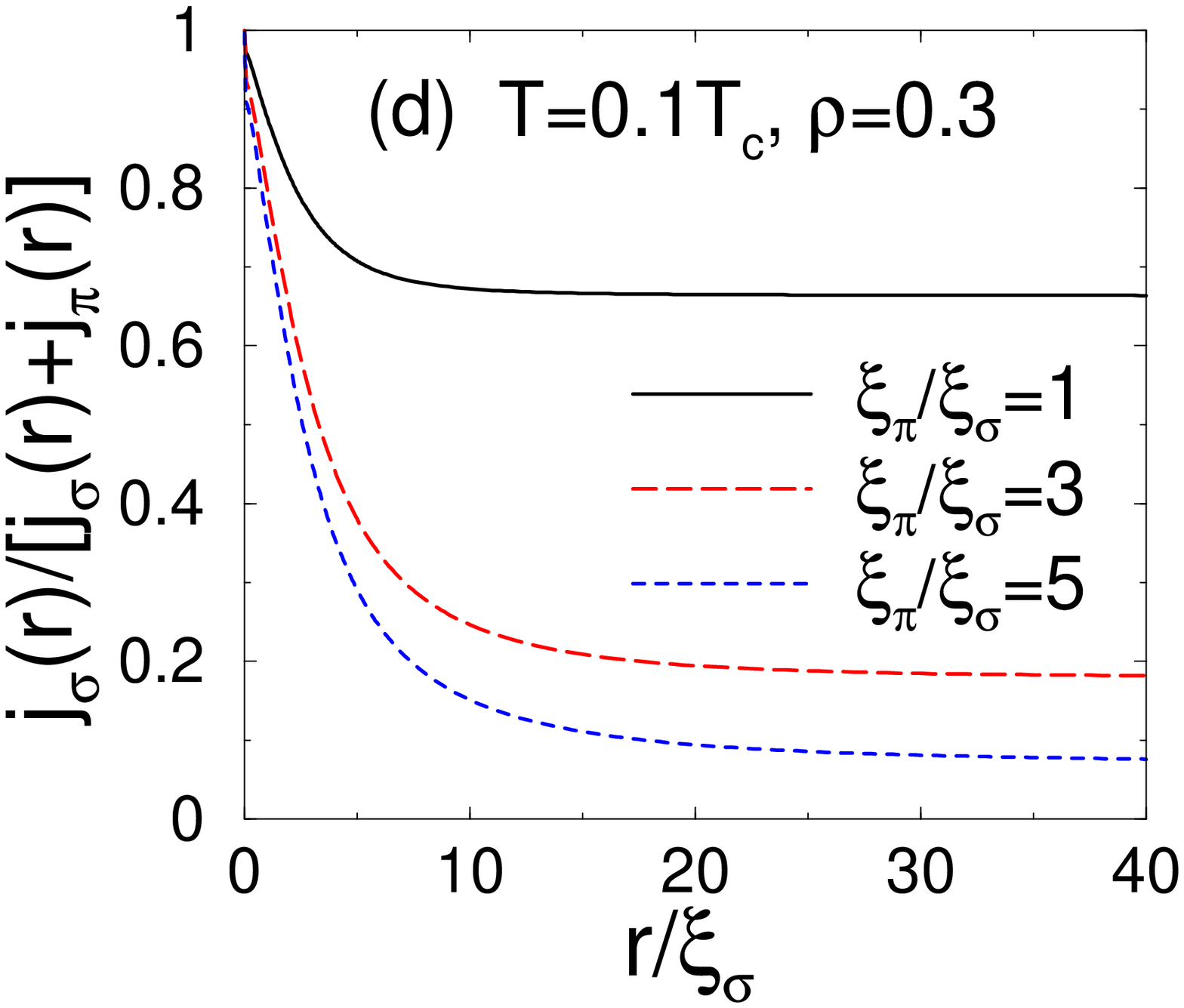}
\end{minipage}
\caption{
Contribution to the
current density as a function of distance from the vortex center,
shown separately for the $\pi$ band and the $\sigma $ band for
fixed mixing ratio $\rho=0.3 $ and different temperatures $T$;
(a) for $\xi_\pi/\xi_\sigma =1$, (b) $\xi_\pi/\xi_\sigma =3$,
and (c) $\xi_\pi/\xi_\sigma =5$.
In (d) the $\sigma $-band contribution to the total current density
is shown as a function of radial coordinate, for
$T=0.1 T_c$, $\rho=0.3 $, and different $\xi_\pi/\xi_\sigma $.
}
\label{fig4}
\end{figure}

The bound-state spectrum is most clearly discussed in terms of the
angle-resolved spectra, shown in Fig.~\ref{fig2} (c). Here the
spectrum of quasiparticles moving in the $x$ direction is shown as a
function of position along the $y$ axis. The position of the bound
states as a function of $y$ is shown in Fig.~\ref{fig2} (d).  The main
bound state branch crosses the chemical potential in the vortex center. The
additional bound states are seen near the gap edge, and show a weak
dispersion. In fact, a close inspection reveals that there are two
additional branches; however, only one of them is 
present at the vortex center.

Finally, in Fig.~\ref{fig4} 
we show contributions from the two bands to the
supercurrent density around the vortex separately.  
We show the results for $\rho=0.3$, and (a) $\xi_\pi/\xi_\sigma =1$, 
(b) $\xi_\pi/\xi_\sigma =3$, and (c) $\xi_\pi/\xi_\sigma =5$, at
$T/T_c=0.1$, 0.3 and 0.5.
The current density from the
$\sigma $ band is enhanced for low temperatures near the vortex
center, and the maximum approaches the center for $T\to 0$.  This is
due to the well-known Kramer-Pesch effect for the clean $\sigma $
band.  
The current density due to the induced superconductivity
in the $\pi$ band is also enhanced by decreasing temperature, but
the maximum does not approach the vortex center as in the $\sigma $ band.  
With increasing $r$, the contribution of the $\sigma $ band is
reduced, and the contribution of the $\pi $ band becomes considerable when 
$\xi_\pi/\xi_\sigma \gtrsim 1$.
At the same time,
${\bf j}_\sigma$ shows temperature dependence
only in the core area, whereas ${\bf j}_\pi $
is temperature dependent also far outside the core.
To discuss this effect, we note that at a large distance $r$
from the vortex center (but small compared to the London
penetration depth), the current-density magnitudes 
are approximately given by
$j_{\sigma }({r}) \sim eN_{F\sigma}  v_{F\sigma }^2/2r $ and
$j_{\pi}({r}) \sim eN_{F\pi} \pi D |\Delta_{\pi }|/r$.
The temperature dependence of $j_\pi$ is thus dominated by that of
$|\Delta_\pi |$.

Another important observation is that, for the parameters appropriate
for MgB$_2$, the contribution of $j_\pi$
is considerable outside the vortex core.
In fact, already for a moderate ratio $\xi_\pi/\xi_\sigma =3$
the $\sigma $-band
contribution is restricted to the region very close to the vortex center 
and is negligible
outside the core for $\rho \ge 0.3$. To understand this effect,
shown in Fig.~\ref{fig4} (d), we note that the current-density ratio
far away from the vortex center approaches
\begin{eqnarray}
\lim_{r\gg \xi_{\pi,\sigma} }\frac{j_{\pi}({r})}{ j_{\sigma}({r})}
=\frac{2D \pi |\Delta_{\pi }|}{v_{F\sigma }^2}
= \frac{|\Delta_{\pi }|}{T_c} \left(\frac{\xi_\pi }{\xi_\sigma }\right)^2
\label{ratio} .
\end{eqnarray}
Clearly, the $\pi $-band contribution to the current density 
dominates when $(\xi_\pi /\xi_\sigma )^2 > (|\Delta_\pi |/T_c)^{-1}$,
which is a case relevant for MgB$_2$. 

In conclusion, we have formulated a model for coupled
ballistic and diffusive bands in terms of coupled Eilenberger and Usadel
equations. We have studied the effects of induced superconductivity
in the `weak' diffusive band on 
the order parameter, the current density and the spectral properties 
of the `strong' ballistic band.
We have found that (a) the recovery lengths of the order parameters in the two
bands are renormalized by Coulomb interactions;
(b) the vortex core spectrum in the $\sigma $ band shows additional bound states
at the gap edges;
and (c) the current density is dominated in the vortex core by the
$\sigma $-band contribution, and outside the vortex core the $\pi $ band
contribution is substantial, or even dominating, for parameters
appropriate for MgB$_2$. 
Our predictions concerning the vortex-core spectrum of the $\sigma $ band
can be tested in future tunnelling experiments.

We acknowledge discussions with
B. Jank\'o, M. Iavarone, M.W. Kwok, and H. Schmidt,
and support by the
NSERC of Canada, the U.S.
DOE, Basic Energy Sciences (W-7405-ENG-36),
the NSF (DMR-0381665),
and the Deutsche Forschungsgemeinschaft within the CFN.

\end{document}